\documentclass[sigconf
]{acmart}

\usepackage{mathptmx}
\usepackage{amsmath,amsfonts,amsthm}
\usepackage{algorithmic}
\usepackage{graphicx}
\usepackage{textcomp}
\usepackage{xcolor}
\usepackage{soul}
\usepackage{fancyhdr}
\usepackage{caption}
\usepackage{comment}
\usepackage{multirow}
\usepackage{enumitem}
\usepackage{todonotes}
\usepackage{subfig}
\usepackage[ruled,vlined,linesnumbered,noend]{algorithm2e}

\usepackage{array}
\usepackage{multirow}
\usepackage{tablefootnote}
\usepackage{comment}
\usepackage{csquotes}
\usepackage{flushend}
\usepackage{float}
\usepackage{footnote}
\makesavenoteenv{tabular}
\makesavenoteenv{table}
\usepackage[normalem]{ulem}

\SetCommentSty{mycommfont}

\begin{document}

\title{HyGain: High Performance, Energy-Efficient Hybrid Gain Cell based Cache Hierarchy}

\author{Sarabjeet Singh}
\authornote{Both authors contributed equally to this research.}
\authornote{This work was carried out while author was at Ashoka University.}
\affiliation{%
  \institution{University of Utah}
  \country{USA}
}
\email{sarab@cs.utah.edu}

\author{Neelam Surana}
\authornotemark[1]
\authornote{This work was carried out while authors were at Indian Institute of Technology, Gandhinagar.}
\affiliation{%
  \institution{Ceremorphic}
  \country{India}
  }
\email{neelam.surana@alumni.iitgn.ac.in}

\author{Pranjali Jain}
\authornotemark[3]
\affiliation{%
  \institution{University of California, Santa Barbara}
  \country{USA}
}
\email{pranjali.jain@alumni.iitgn.ac.in}

\author{Joycee Mekie}
\affiliation{%
  \institution{Department of Electrical Engineering, Indian Institute of Technology, Gandhinagar}
  \country{India}
}
\email{joycee@iitgn.ac.in}

\author{Manu Awasthi}
\affiliation{%
  \institution{Ashoka University}
  \country{India}
}
\email{manu.awasthi@ashoka.edu.ac.in}

\renewcommand{\shortauthors}{Singh and Surana, et al.}

\begin{abstract}
In this paper, we propose a ``full-stack'' solution to designing high capacity and low latency on-chip cache hierarchies by starting at the circuit level of the hardware design stack. First, we propose a novel Gain Cell (GC) design using FDSOI. The GC has several desirable characteristics, including \textasciitilde 50\% higher storage density and \textasciitilde 50\% lower dynamic energy as compared to the traditional 6T SRAM, even after accounting for peripheral circuit overheads. We also exploit back-gate bias to increase retention time to 1.12 ms (\textasciitilde 60$\times$ of eDRAM) which, combined with optimizations like staggered refresh, makes it an ideal candidate to architect all levels of on-chip caches. We show that compared to 6T SRAM, for a given area budget, GC based caches, on average, provide 29\% and 36\% increase in IPC for single- and multi-programmed workloads, respectively on contemporary workloads including SPEC CPU 2017. We also observe dynamic energy savings of 42\% and 34\% for single- and multi-programmed workloads, respectively.  

We utilize the inherent properties of the proposed GC, including decoupled read and write bitlines to devise optimizations to save precharge energy and architect GC caches with better energy and performance characteristics. Finally, in a quest to utilize the best of all worlds, we combine GC with STT-RAM to create hybrid hierarchies. We show that a hybrid hierarchy with GC caches at L1 and L2, and an LLC split between GC and STT-RAM, with asymmetric write optimization enabled, is able to provide a 54\% benefit in energy-delay product (EDP) as compared to an all-SRAM design, and 13\% as compared to an all-GC cache hierarchy, averaged across multi-programmed workloads.

\end{abstract}


\maketitle

\section{Introduction}
\label{section:introduction}

With the~\cite{Lamport:LaTeX} increasing number of cores on-chip~\cite{CacheDesignforCMP},
 additional memory is needed to 
feed these cores. Emerging workloads have become significantly memory 
intensive and have large working set sizes~\cite{MemoryBehaviourofWorkloads,sarabjeet2019_SPEC}. These factors have necessitated the need for high capacity, low latency, on-chip caches.

Several research efforts have been made to increase capacity 
and contain latency of on-chip caches, which has led to ever-increasing cache capacities and deeper cache hierarchies~\cite{MemoryBehaviourofWorkloads}. However, the memory technology,
which makes up the bulk of on-chip caches, has remained unchanged. On-chip caches, at almost all levels of the memory 
hierarchy, have been devised using 6T SRAM. Even though SRAM 
suffers from low areal density, high leakage power and 
high dynamic energy requirements compared to other memory technologies~\cite{L3C_EmergingTechinCache,weste,rabaey}, the latency superiority 
 of SRAM, and its compatibility with logic fabrication technology 
 has made it indispensable for creating low-latency caches.

Recently, a number of alternative memory technologies have started 
to emerge, and have been evaluated for use in caches. Contenders for SRAM replacement include non-volatile memory technologies like Spin-Transfer Torque RAM 
(STT-RAM)~\cite{STTRAM_Cache_LLC,STTRAM_Cache_3D,STTRAM_Cache_TradeoffwithNonVolatility} and 
Phase Change Memory (PCM)~\cite{PCM_Cache_Hybrid}, as well as volatile 
ones like embedded DRAM (eDRAM)~\cite{L3C_EmergingTechinCache}. 

In addition to providing data non-volatility, both STT-RAM and PCM provide 
3-4$\times$ density benefits over 6T SRAM~\cite{STTRAM_Cache_3D,PCM_Cache_Hybrid}, making them attractive candidates for high 
capacity caches. However, there are drawbacks inherent to both technologies, 
including higher write energy (up to \textasciitilde 5$\times$ that of SRAM) and access latencies (\textasciitilde 1.5$\times$ read, 5$\times$ write latency for STT-RAM, PCM is worse)~\cite{STTRAM_Cache_3D,PCM_Cache_Hybrid}. In most cases, the drawbacks
outweigh the benefits, rendering these technologies suitable for use only in 
last-level caches, where both capacities and access latencies are 
expected to be higher.
\begin{figure}
  \centering
  \includegraphics[width=\linewidth]{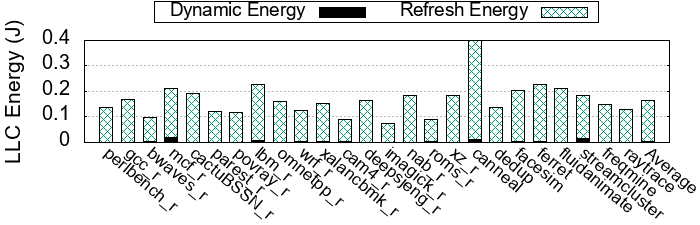}
  \caption{eDRAM LLC energy breakdown.}
  \label{fig:eDRAM_LLC}
\end{figure}
\begin{table*}
    \small
    \caption{Comparison of various memory technologies for on-die caches. Limitations of each technology in bold.
    }
    \label{table:technologycomparison}
    \begin{center}
    \begin{tabular}{|c||c|c|c|c|c|c|c|c|c|}
    \hline
     & SRAM & STT-RAM \cite{STTRAM1} & PCM\cite{PCM_Cache_Hybrid} & 1T1C eDRAM\cite{embeddedcache} & 2T GC\cite{2Tcache} & 3T GC \cite{3Tcache} & 4T GC \cite{4Tcache} & \textbf{Proposed} \\
    \hline
    Total Transistors & 6T & 1T+1MTJ & 1T+1PCM & 1T+1C & 2T & 3T & 4T & 2T\\
    \hline
    Area (um$^2$) & \textbf{0.64} & 0.16 & 0.16 & 0.16 & 0.24 & 0.38 & \textbf{0.54} & 0.24 \\
    \hline
    Data Storage & Latch & Magnetization & Phase & Capacitor & MOS & MOS & MOS & MOS \\
    \hline
    Read/Write Time & Short/Short & Short/\textbf{Long} & Short/\textbf{Long} & Short/Short & Short/Short & Short/Short & Short/\textbf{Long} & Short/Short  \\
    \hline
    Read/Write Energy & X/Y & X/\textbf{5Y} & X/\textbf{5Y} & 0.5X/0.5Y & 0.5X/0.5Y & 0.5X/0.5Y & 0.5X/0.5Y & 0.5X/0.5Y  \\
    \hline
    Leakage/Yield & \textbf{High}/High & Low/High & Low/High & Low/\textbf{Low} & Low/High & Low/High & Low/High & Low/High \\
    \hline
    Retention Time & - & - & - & \textbf{20 us} & \textbf{20 us} & \textbf{20 us} & 1.6 ms & 1.12 ms  \\
    \hline
    \begin{tabular}[c]{@{}l@{}}Destructive Read/\\ Decoupled Bitline\end{tabular} & No/\textbf{No} & No/Yes & No/Yes & \textbf{Yes}/\textbf{No} & No/Yes & No/Yes & No/Yes & No/Yes \\
    \hline
    \end{tabular}
    \end{center}
    \normalsize
\end{table*}
eDRAM has also been evaluated as a candidate for architecting caches~\cite{L3C_EmergingTechinCache,ReducingCachePowerwithECC}. It has found 
 adoption in multiple recent products, including IBM's Power series~\cite{IBM}, Intel's Haswell~\cite{INTEL} and Microsoft's
Xbox 360~\cite{xbox}, again as a technology for LLCs. 
Since eDRAM is a DRAM variant, it provides higher 
density compared to SRAM and has favorable access latency profiles~\cite{L3C_EmergingTechinCache} as compared to NVMs. However, the traditional 
1T1C eDRAM cells suffer from low Data 
Retention Times (DRTs) of 20 - 50 $\mu$s~\cite{L3C_EmergingTechinCache}, requiring frequent refreshes in many eDRAM based design. 
These refreshes cause significant energy consumption as the data from 
an eDRAM row has to be read and written back~\cite{L3C_EmergingTechinCache,ReducingCachePowerwithECC}, making addressing refresh operations as the primary challenge in designing eDRAM caches. 
The energy consumption breakdown of dynamic and refresh energies for single programmed SPEC CPU2017 and PARSEC workloads in an eDRAM LLC (simulation 
parameters are listed in Section~\ref{section:TechComparison}) is shown in 
Figure~\ref{fig:eDRAM_LLC}. As can be observed, refresh energy is many times 
higher than dynamic energy, which makes optimizing refresh operations
 an essential consideration for designing eDRAM based caches.

Apart from energy overheads, an eDRAM row, and hence a portion of the 
 cache, is unavailable for the duration of the refresh period, leading to
 performance overheads. To counter the shortcomings of traditional eDRAM, 
 another eDRAM variant, Gain Cell (GC) has been 
 proposed~\cite{2Tcache,3Tcache}, which has multiple advantages over 
  1T1C eDRAM. These include a cheaper, logic-compatible fabrication 
  process and the absence of a dedicated capacitor per 
  cell; GCs use the transistor's parasitic capacitance for data storage. 
  Finally, GCs provide non-destructive reads~\cite{L3C_EmergingTechinCache}
  and decoupled read/write bitlines, resulting in lower access latency and energy consumption~\cite{2Tcache}.

Despite these advantages, traditional 2T and 3T GCs have not found 
adoption widespread since they suffer from low DRTs, hence requiring frequent 
refreshes. While 4T GCs with higher DRTs have been 
proposed~\cite{4Tcache}, they suffer from higher write energies 
and lower density, making them unattractive as 1T1C eDRAM replacements.
In any case, the presence of refresh makes existing GCs 
useless at any level of cache, other than LLCs \cite{L3C_EmergingTechinCache}.

As a result, even though each one of these technologies has its 
pros and cons, they cannot be used as a \emph{drop-in} replacement for SRAM 
caches, especially for levels closer to the CPU. However, in the 
presence of a suitable SRAM replacement, many of their pros can 
be combined to architect high capacity caches that have similar 
latency profiles as that of an SRAM based hierarchy.
In this paper, we attempt such a design, starting from ground up.
First, we propose a novel FDSOI MOSFET~\cite{FDX} based \textit{2T} Gain
Cell, which provides high storage density, low access latencies, and a high 
DRT. This makes the cell amenable for use \textit{at all levels of the cache 
hierarchy}. Even after accounting for peripheral circuitry overheads, compared 
to 6T SRAM, the proposed GC array offers a 50\% reduction in read/write 
energies, 50\% reduction in the area while keeping access latency unchanged. It 
exhibits  low leakage energy and has modest refresh requirements. A comparison 
of advantages of the proposed GC over competing memory technologies for caches is presented in Table~\ref{table:technologycomparison}. The main contributions of this work are summarized below:
\begin{itemize}
\itemsep-0.01em 
    \item We propose a novel, FDSOI based 2T Gain Cell, specifically for use in on-chip caches, and exploit its back-gate bias feature to reduce leakage power by 99\% and increase retention time by \textasciitilde 60$\times$, compared to conventional GC and eDRAM, reducing the need for frequent refreshes. We combine this already large refresh window with
    optimizations like staggered refresh, to design practically refresh-free GC caches. As a result, proposed GCs can be used at \textit{all} levels 
    of the cache hierarchy.
    \item We show that GC based sub-arrays exhibit 2$\times$ area advantage
    and similar latency characteristics as compared to SRAM, even after accounting for overheads of peripheral
    circuits. This helps architect higher capacity caches
     within the same area and 
    latency budgets. As a result, for single-programmed workloads, iso-area caches architected using proposed GCs at all levels of the hierarchy 
    exhibit a 42\% reduction in dynamic energy and a 29\% increase in IPC as compared to SRAM. Multi-programmed workloads exhibit similar behavior.
    We further show that the proposed GC scales 
    well at smaller technology nodes, and retains its advantages over SRAM.
    \item We utilize the inherent decoupling in the read and write bitlines in  proposed GCs to save precharge energy between consecutive writes, and clubbing up to 70\% writes with reads, thereby improving performance and reducing dynamic energy consumption by 13\%, as compared to GC based caches lacking this optimization. We also explore 
    optimizations like no-refresh policy, where a line is invalidated if 
    it is not accessed during its DRT period, and show that this can be 
    used as an effective mechanism to eliminate refreshes in caches closer to 
    the CPU \textit{without} performance penalties.
    \item Finally, to create high capacity, low latency caches, 
    we evaluate several hybrid cache hierarchies by incorporating emerging technologies like STT-RAM with GCs to 
    design caches that can provide up to 4$\times$ higher capacity, compared 
    to iso-area SRAM caches. For multi-core workloads, hybrid caches architected with GCs, combined with asymmetric writes optimization,
    can provide 43\% performance and 44\%
    energy  
    benefits as compared to iso-area SRAM caches.
  \end{itemize}
The remainder of this paper is organized as follows. We provide the 
circuit level implementation details for proposed GC in
Section~\ref{sec:background} and details of the architectural implementations
of GC caches in Section~\ref{section:GainCellCaches}.
Section~\ref{section:experimental_methodology} presents the experimental 
evaluation setup, while Section~\ref{section:experimental_results} analyses the energy and performance implications of the implementations. Section~\ref{section:asymmetricwrites} proposes optimizations by exploiting intrinsic properties of GC, while Section~\ref{section:TechComparison} evaluates 
hybrid cache hierarchies. Sections~\ref{section:ScalingGC} discusses the scalability of GC at lower technology nodes and provides an assessment of the 
no-refresh policy. Finally, we discuss related work in Section~\ref{section:relatedwork} and conclude in Section~\ref{section:conclusion}.
\begin{figure}
  \begin{center}
  \includegraphics[width=\linewidth]{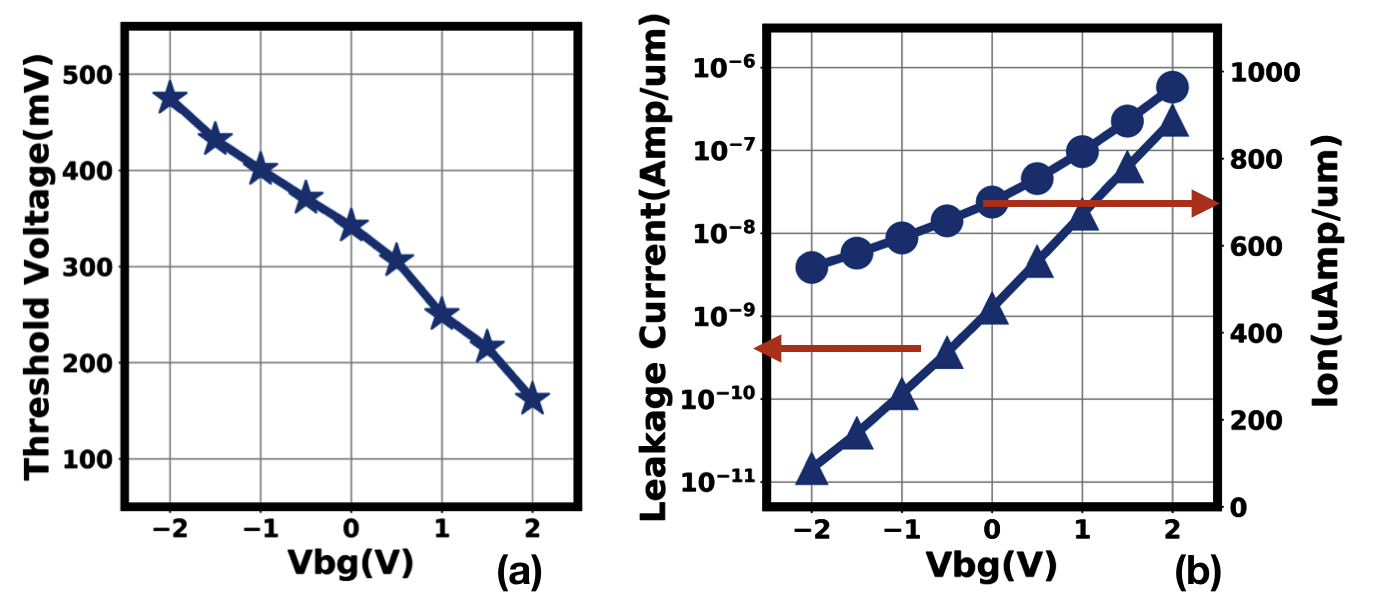}
   \caption{The effect of back-gate bias voltage on (a) Threshold Voltage (V$_{TH}$) (b) Leakage Current (I$_{OFF}$) and On-state (I$_{ON}$) in n-type FDSOI transistor.}
   \label{fig:Id_Vgs_Backgate_characterics}
  \end{center}
\end{figure}
\begin{figure}
\centering     
\subfloat[Proposed 2T GC]{\includegraphics[width=0.4\linewidth]{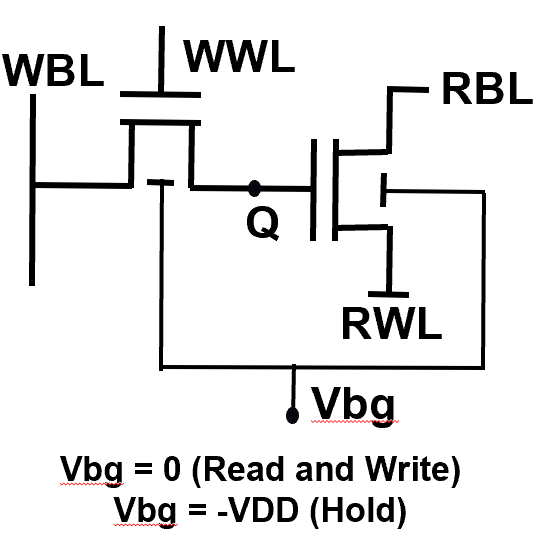}\label{fig:schematic_GC}}
\subfloat[6T SRAM Cell.]{\includegraphics[width=0.4\linewidth]{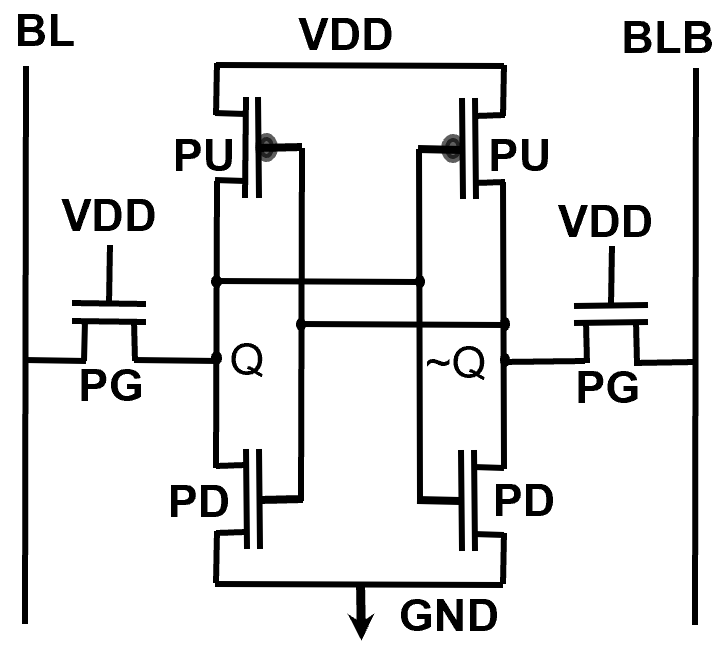}\label{fig:schematic_SRAM}}
\caption{Schematic diagram of the memory cell. }
\label{fig:cell_schematic}
\end{figure}

\section{Back-Gate Biased Gain Cell}
\label{sec:background}
Gain Cells have started gaining traction owing to their logic-compatible fabrication process, small area footprint, and low energy requirements~\cite{3Tcache,4Tcache,gaincell3,2Tcache}. GCs have
been fabricated and tested in FinFET~\cite{finfetgain},
bulk~\cite{capacitorless}, and FDSOI~\cite{4Tcache} processes. However, these proposals still suffer from low DRT leading to enormous
refresh energy, as shown in Table~\ref{table:technologycomparison}. Earlier, FDSOI
devices have been fabricated in 12 nm~\cite{Global2} and are
expected to scale down, considering their advantages
of higher DRT over FinFET\cite{fdsoifinfetprocess,finfet+fdsoi}.
We implement an n-type 2T GC on an
FDSOI device~\cite{fdsoi1,FDX} and exploit its back-gate  bias feature to lower the leakage current, thus improving the DRT. 

\subsection{Using Back-gate Biased FDSOI to increase DRT} 
\label{subsec:fdsoi}
Data retention time (DRT) of a transistor is directly linked with the leakage current of the transistor in the OFF condition. The leakage current ($I_{OFF}$), in turn, exponentially depends on the threshold voltage~($V_{TH}$) of the transistor. In the recent past, with technology scaling, leakage has prohibitively increased in bulk-MOSFETs necessitating significant manufacturing efforts and additional fabrication steps to control the leakage current. Silicon-on-Insulator (SOI) technology offers a promising solution to deal with leakage current due to an additional handle to control the threshold voltage using its back-gate biasing. The junction leakage currents are significantly reduced in a fully-depleted SOI (FDSOI) as an oxide layer removes the p-n junction from the substrate, and has been used in a few commercial offerings, including IBM's POWER 8~\cite{POWER8}. Due to this oxide layer, the substrate works as a second gate or the back-gate\cite{Global}, and can be biased to change the threshold voltage of the FDSOI transistor which, in-turn, exponentially reduce the leakage current~\cite{powergating,multiplier,back-gate1,back-gate2,back-gate3,back-gate4},  and increases the DRT of the transistor. 

We have implemented the n-type  transistor using ST Microelectronics 28-nm FDSOI technology and simulated it using Cadence Virtuoso.  Figure~\ref{fig:Id_Vgs_Backgate_characterics}(a) captures the effect of back-gate biasing ($V_{bg}$) on threshold voltage, when $V_{bg}$ is varied between $-2V$ to $+2V$.  Figure~\ref{fig:Id_Vgs_Backgate_characterics}(b) shows the exponential dependence of $I_{OFF}$ on $V_{bg}$, and as $V_{bg}$ becomes more negative, the leakage current exponentially drops by almost four orders of magnitude, up
to 10 pA/$\mu$m. Figure~\ref{fig:Id_Vgs_Backgate_characterics}(b) also shows the improvement on $I_{ON}$ (on-current) when $V_{bg}$ is positive. Improvement in $I_{ON}$ makes the transistor operate faster, thus reducing delay. Thus, back-gate biasing can be used both to improve performance (latency) and reduce leakage. The schematic of the proposed FDSOI based Gain Cell is shown in Figure~\ref{fig:schematic_GC}. During hold
condition, -V$_{DD}$ is applied to the back-gate bias of the device,  which reduces the leakage from W1, improving DRT.
\subsection{Overheads of back-gate biasing}
Leveraging back-gate bias for DRT improvement leads to the reduction of ON current of n-type FDSOI, as depicted in Figure~\ref{fig:Id_Vgs_Backgate_characterics}(b), and hence, higher cell access latency.
To capture the best of both worlds, we apply zero back-gate bias during read and write operations, and -V$_{DD}$ during the hold condition. This can be implemented with the circuit proposed in~\cite{back-voltage}.

\begin{figure}[h]
 \begin{center}
  \includegraphics[width=\linewidth]{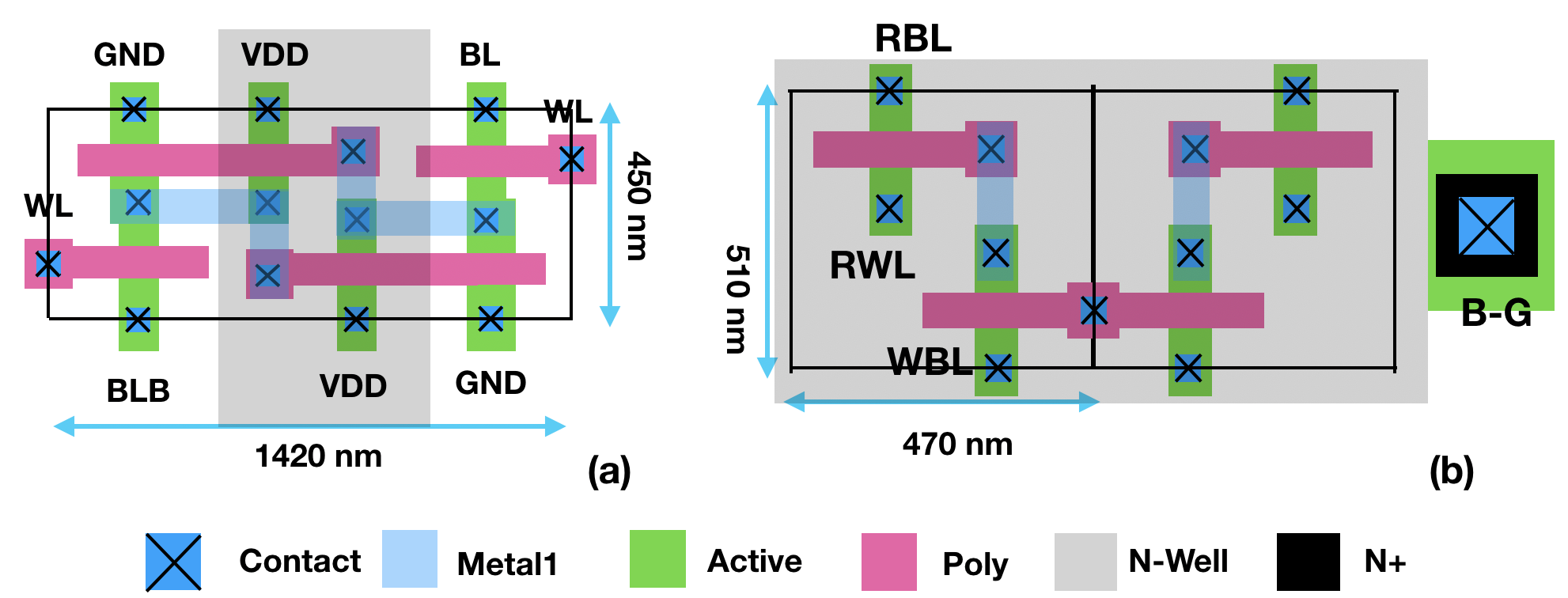}
   \caption{Layout of (a) 6T SRAM Cell (b) Proposed 2T Gain Cell (all cells in a row share a single back-gate, layout shows 2 GCs with shared B-G).}\label{fig:cell_layout}
  \end{center}
\end{figure}

Figure~\ref{fig:cell_layout} shows the layouts of SRAM and proposed 2T GC. To implement back-gate bias, a single contact can be shared across all the cells of a row in an array (as shown in Figure~\ref{fig:cell_layout}(b)). This keeps the area overhead minimal. Since the back-gate oxide thickness is quite large (\textasciitilde 20 nm), back-gate carries only 5\% of the front-gate
capacitance, which keeps switching power overhead less than 5\%, while increasing DRT. Additionally, since V$_{bg}$ and row-signal (WWL) switching happens in parallel, delay of back-gate is over-shadowed by WWL's delay. Hence, back-gate bias causes \textit{no} latency penalty, negligible area penalty, and has only \textasciitilde 5\% of the switching power overhead.

\begin{figure}[h]
  \begin{center}
  \includegraphics[width=\linewidth]{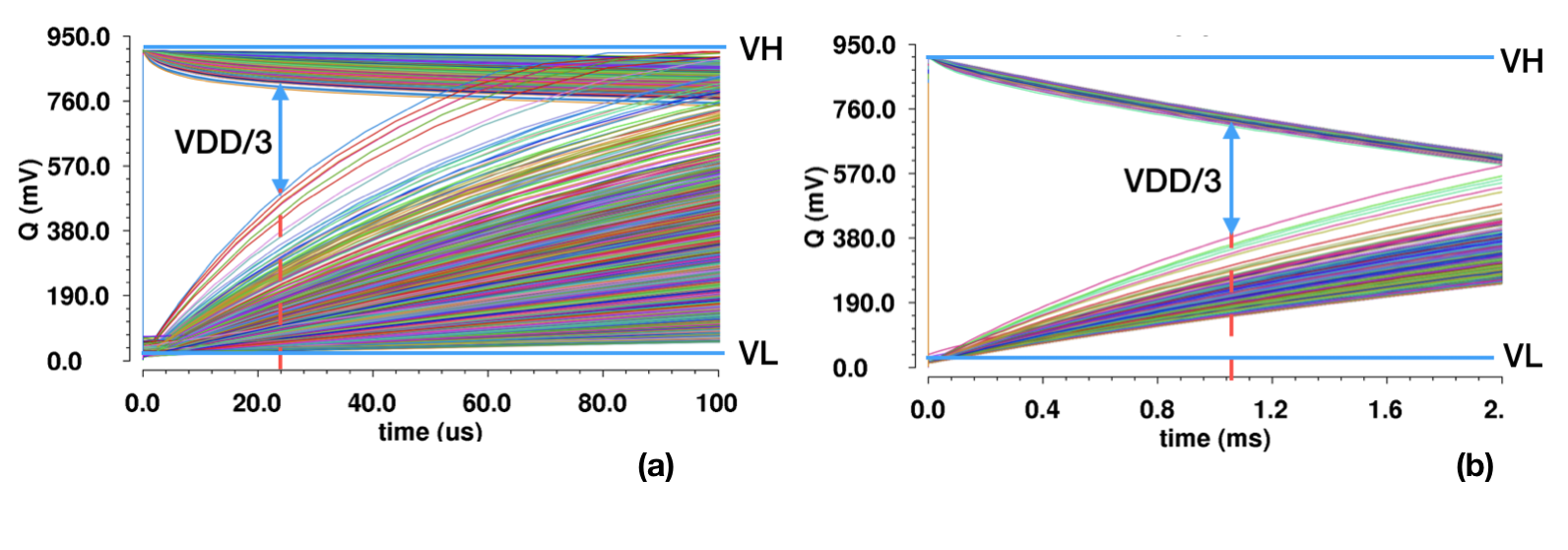}
   \caption{10K, Monte-Carlo waveforms of 1 and 0 decay (a) Traditional 2T Gain Cell (b) Proposed 2T Gain Cell.}
   \label{fig:retention_time}
  \end{center}
\end{figure}
\begin{figure}[h]
  \begin{center}
  \includegraphics[width=\linewidth]{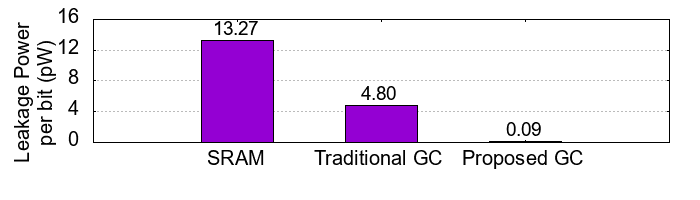}
   \caption{Leakage Power of 6T SRAM, Traditional GC \& Proposed GC}\label{fig:leakage_power_comparison}
  \end{center}
\end{figure}

\subsection{Working of Back-Gate Biased Gain Cell (BGB-GC) }
Figure~\ref{fig:cell_schematic}(a) shows the schematic of the proposed
n-type 2T GC. GC has decoupled read and write operations
and has non-destructive reads, unlike 1T1C eDRAM. The input
signals for these operations are illustrated in Table~\ref{table:celloperations}. Working of proposed 2T GC is similar to conventional 2T GC, except that we use back-gate bias during hold condition.    
\subsubsection{Write Operation} \label{subsubsec:write_operation}
Write Bitline (WBL) is shared across an entire column in an array and has a large capacitance. Write Wordline (WWL) runs along the row in the array. To write to the cell, data is first transferred to WBL, and then a row-signal (WWL) is used to transfer data to the Q node. 

\begin{table}
\caption{Working of the Proposed Gain Cell}
\label{table:celloperations}
\scriptsize
\begin{center}
\begin{tabular}{|c|c|c|c|c|c|}
\hline
\textbf{Operation} & \textbf{WBL} & \textbf{WWL} & \textbf{Back-gate(bg)} & \textbf{RBL} & \textbf{RWL} \\
\hline
\textbf{Read}  & $-$ & 0 & 0 & V$_{DD}$(floating) & 0 \\
\hline
\textbf{Write} & Data & V$_{DD}$ & 0 & V$_{DD}$ & V$_{DD}$ \\
\hline
\textbf{Hold} & $-$ & 0 & -V$_{DD}$ & V$_{DD}$ & V$_{DD}$ \\
\hline
\end{tabular}
\end{center}
\normalsize
\end{table}  

\subsubsection{Read Operation} \label{subsubsec:read_operation}
For a read operation, Read Bitline (RBL), which is a shared
signal across the column, is first precharged to V$_{DD}$. Then,
to read data, RBL is kept floating. Active low signal to Read Wordline (RWL) is used to read the data. If data stored in Q is 1, RBL discharges; otherwise it remains at V$_{DD}$, which is sensed by a sense amplifier. During the read
operation, energy is consumed in the switching of RBL and
RWL. Most importantly, read operation in GC is non-destructive, since RBL is
decoupled from the Q node~\cite{2Tcache}.

\subsubsection{Hold Condition} \label{subsubsec:hold_condition}
Periods where the cell is neither read nor written to is known as the hold condition. This is important since the cell still needs to retain data during this period, unlike SRAM where data is retained due to cross-coupled inverters shown in Figure~\ref{fig:schematic_SRAM}. The charge in Q node leaks from W1 over time, necessitating refresh operations for data restoration, before the DRT window closes.

\subsection{BGB-GC comparison with 6T SRAM}
\subsubsection{Data Retention Time (DRT)}
The proposed GC uses back-gate bias voltage of -V$_{DD}$ during hold operation, leading to reduction in leakage current from the W1 transistor, improving DRT. To quantify DRTs, we performed Monte Carlo simulations considering the standard 6-$\sigma$ local and 1-$\sigma$ global process variations. Figure~\ref{fig:retention_time} shows the data degradation of conventional and proposed 2T GC, for 10K M-C simulations. DRT is measured at a point where the data can be read without error. Typically, V$_{DD}$/3 is a sufficient margin to read data properly, and we consider the worst-case DRT as the refresh interval, making these results more pessimistic than usual. For conventional 2T GC~\cite{2Tcache}, DRT obtained is 19$\mu$s by considering 100\% yield (Figure~\ref{fig:retention_time}(a)), which is consistent with~\cite{4Tcache}. For the proposed GC, the data decay has
significantly slowed down, as seen from
Figure~\ref{fig:retention_time}(b). We note that the worst-case DRT obtained for the proposed GC \textit{improves to 1.12 ms}, which is \textbf{\textasciitilde 60$\times$} higher than the case when no back-gate bias is applied. This is \emph{first} such GC proposal built with 2 transistors leading to significant gains in DRT over existing GC designs, except 4T GC designs~\cite{4Tcache} where a similar DRT is achieved at the cost of increased area and latency.



\subsubsection{Leakage Power}
Since the proposed 2T GC has a smaller number of leakage paths compared to SRAM
(schematic in Figure~\ref{fig:cell_schematic}), it inherently has
lower leakage power. Additionally, we have used
back-gate bias to further reduce the leakage current significantly. We compare the leakage power of 6T SRAM, GC without back-gate bias, and proposed GC with back-gate bias in
Figure~\ref{fig:leakage_power_comparison}. We show that proposed
GC has \textbf{\textasciitilde 99\%} reduction in leakage power as compared to 6T SRAM.

\subsubsection{Area}
Figure~\ref{fig:cell_layout} compares the layout of the 6T SRAM cell and 2T GC at 28 nm technology. SRAM cell takes 0.64 $\mu$m$^2$, whereas the proposed GC takes 0.24 $\mu$m$^2$, which is 40\% of the SRAM cell. The smaller size of the proposed GC allows for much higher density for the same area.

At the cell level, the proposed GC takes only 0.4$\times$ area compared to the 6T SRAM cell. Layout of the SRAM cell is drawn in a very efficient way and have area efficiency of 80\%-90\%\cite{area1,weste,rabaey}. Considering this, at the cache level, GC can have \textasciitilde 2$\times$ capacity as compared to SRAM cache. 
Even though GC has 2.5$\times$ benefits at the cell level, at the cache level, it reduces to 2.0$\times$ due to peripheral circuitry overhead.
In the rest of the paper, for the iso-area comparison, we have considered 2$\times$ capacity of GC compared to the SRAM.       

\section{Architecting Gain Cells for Caches}
\label{section:GainCellCaches}

First, we describe the sub-array construction 
of proposed GCs, and the mechanism by which it maps to 
various cache configurations. Then we compare the architectural 
benefits of GCs over SRAM.

\begin{figure}[h]
  \centering
  \includegraphics[width=\linewidth]{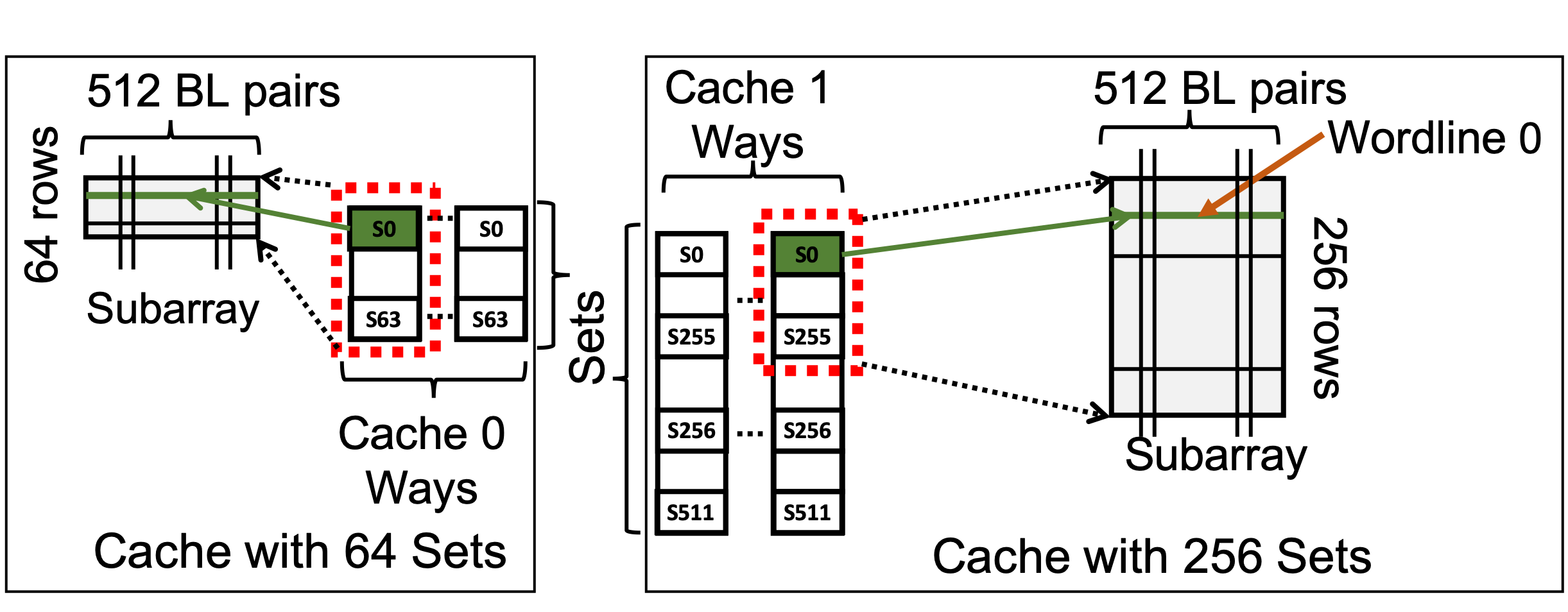}
  \caption{Mapping of Caches to GC Subarrays for different caches}
  \label{fig:subarray-mapping-all}
\end{figure}

GCs are arranged as sub-arrays, in a typical row-column fashion. A cache 
can then be mapped to multiple sub-arrays, as dictated 
by its capacity. From an extensive design space exploration,
 we conclude that the sweet spot for minimum latency and peripheral
 circuitry overheads lie at a sub-array size of 256$\times$512
 bits, or 16~KB. Hence, GC caches can be architected such that 
 each way, across all cache sets, maps to one sub-array. As a result, 
 looking up a cacheline (64B) is the same as looking up a row of this
 sub-array, as shown in the Set0-Way0 to Row0 mappings in Figure~\ref{fig:subarray-mapping-all}. In cases 
 where combined size of a way is $>$16~KB, we keep adding
 sub-arrays, until all the sets have been accounted for. For example,
 the right hand side of Figure~\ref{fig:subarray-mapping-all} depicts 
 a cache where one way is mapped to two 256$\times$512 sub-arrays.
 
 However,  this prohibits mapping of any cache configuration where the combined capacity for one way is smaller than 16~KB, as illustrated in the left half of Figure~\ref{fig:subarray-mapping-all}. For these caches, 
 we keep the design choice 
 of mapping an entire way to one sub-array, while reducing the size of the sub-array. For example, in the case of a 64KB, 16-way cache, an entire way (4KB) is mapped to a 64$\times$512 bit sub-array. This ensures that a 64~B 
 cacheline lookup is not spread across multiple sub-arrays.

\begin{figure}[h]
  \centering
  \includegraphics[width=\linewidth]{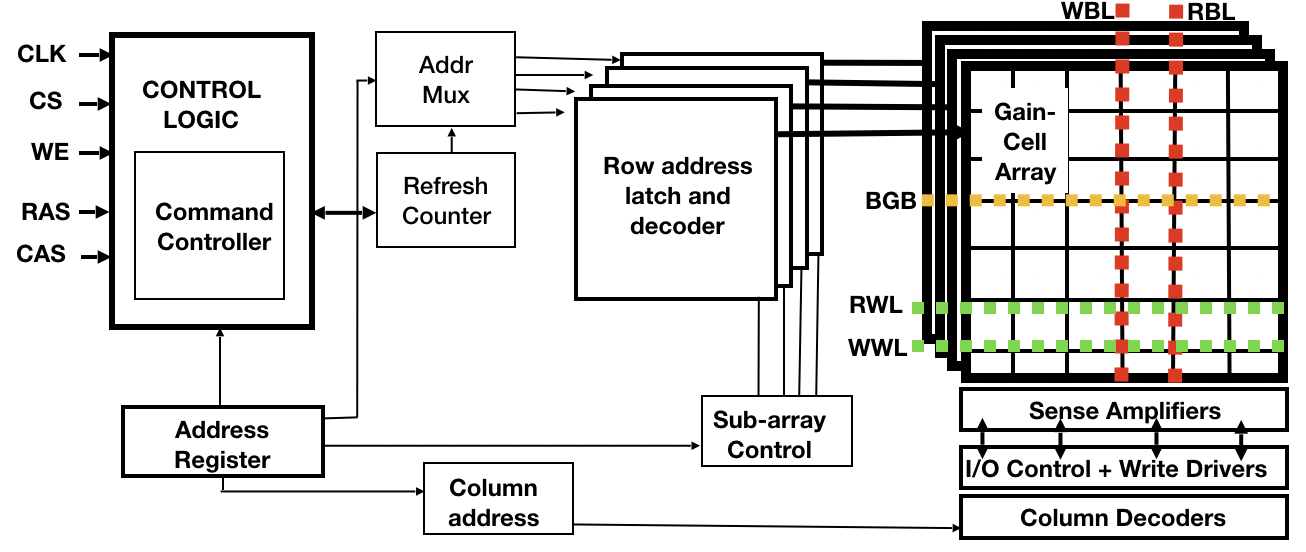}
  \caption{Block diagram of the Proposed Gain Cell based Cache}
  \label{fig:Gain-CellCache}
\end{figure}

\begin{table}
\caption{Cache Characteristics - SRAM v/s Proposed GC}
\label{table:cachetiming}
\centering
\scriptsize
\begin{tabular}{|c|c|c|c|c|}
\hline
\multicolumn{2}{|c|}{\textbf{Cache Level}} & \textbf{32kB L1} & \textbf{256kB L2} & \textbf{8MB L3} \\ \hline
\multirow{2}{*}{Latency (ns) \textbf{(in cycles)}} & SRAM & 0.475\textbf{(2)} & 1.34\textbf{(5)} & 2.81\textbf{(10)} \\ \cline{2-5} 
 & GC & 0.42\textbf{(2)} & 1.20\textbf{(5)} & 2.55\textbf{(9)} \\ \hline
 \multirow{2}{*}{\begin{tabular}[c]{@{}c@{}}Read/Write Energy \\per bit (pJ)\end{tabular}} & SRAM & 0.75/1.13 & 2.18/3.1 & 7.5/11.8 \\ \cline{2-5} 
 & GC & 0.41/0.68 & 1.15/1.63 & 4.1/5.81 \\ \hline
\multirow{2}{*}{\begin{tabular}[c]{@{}c@{}}Write Energy/bit for Same\\ Bit (0-$>$0 or 1-$>$1) (pJ)\end{tabular}} & SRAM & \multicolumn{3}{c|}{Same as Write Energy} \\ \cline{2-5} 
 & GC & 0.24 & 0.7 & 1.95 \\ \hline
\multicolumn{2}{|c|}{Leakage/bit (pW) (SRAM/GC)} & \multicolumn{3}{c|}{13.27/0.09} \\ \hline
\multicolumn{2}{|c|}{Refresh Interval(ms)/ Period per line(ns)} & \multicolumn{3}{c|}{1.12/1.5} \\ \hline
\multicolumn{2}{|c|}{Refresh Energy/bit (pJ)} & \multicolumn{3}{c|}{1.87} \\ \hline
\end{tabular}
\normalsize
\end{table}

Figure~\ref{fig:Gain-CellCache} shows the block diagram of the proposed GC cache. Compared to SRAM cache, GC has additional refresh counters at each level of cache. As per concurrent refresh, the refresh counter will generate signal to refresh the same row in all subarrays at the same time. BGB signal is generated along with RWL and WWL signals while performing refresh (no extra delay penalty). For staggering the refresh across different rows, the counter times out after every DRT/N time (N is number of rows in subarray). 

Next, we compare and contrast the architecture level characteristics of 
on-chip caches devised using SRAM and GCs. 
We extract SRAM and GC energy and
latency parameters using an enhanced CACTI~\cite{cacti6.0} model and 
present these results in Table~\ref{table:cachetiming}. These
results were also validated using SPICE simulations
using ST-microelectronics 28-nm FDSOI CMOS technology.  
We observe that for every cache level, the dynamic read and write energies for a GC cache are reduced by \textit{at least} 50\%, as compared to an SRAM one. This is because the proposed GC requires just 
one bitline per read or write access, thereby reducing a significant fraction of the dynamic energy consumed in switching of bitlines and word lines~\cite{tiered,tiered1}.

\begin{figure}[h]
  \centering
  \includegraphics[width=\linewidth]{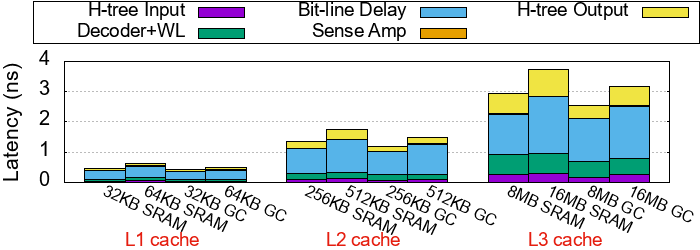}
   \caption{Latency comparison of proposed GC and conventional SRAM caches }\label{fig:latency}
\end{figure}

Additionally, owing to decoupled read and write like 8-T SRAM\cite{rabaey}
operations, GC has slightly lower access latencies as
compared to SRAM, allowing for similar cycle time access
as that of a SRAM cache, for a given processor frequency.
Another trait of GC is high
density, which enables us to fit a similar capacity cache  in \textit{half the area}. Alternatively, in a
given area budget, we can implement a higher capacity
cache by increasing associativity.
As shown in Figure~\ref{fig:latency},
iso-area access latencies of GC based caches at all levels of the 
hierarchy are similar to SRAM caches, with additional benefit of GC caches having twice the capacity. Doubling the capacity
of an SRAM based cache 
increases the area by 2.0$\times$. Not only that, it 
also increases access latency of caches by at least 30\%. 
As a result, GC based caches allow for twice the capacity in 
the same latency \textbf{and} area budget, for every level of cache. 


\subsection{GC Based Cache Proposals}
Using these observations, we propose the use of GCs at
various levels of caches, from all on-chip caches architected using GCs (\textbf{ALL-GC}) to just last-level cache (\textbf{LLC-GC}) or L1 cache (\textbf{L1-GC}) being GC, and compare with the baseline case where all caches are implemented with SRAM (\textbf{ALL-SRAM}). 
Since GCs provide excellent density benefits over SRAM, we examine the energy and performance implications of GC over SRAM for both iso (cache) capacity and iso-area. For iso-capacity
(\textbf{-CAP}), SRAM and GC caches are compared with
the same cache size, which indicates lower on-chip area usage by GC caches. 
While in the case of iso-area (\textbf{-AREA}), the GC caches are doubled in 
capacity by increasing their associativity, while retaining latency 
characteristics. We maintain the tag array in SRAM; only the data arrays 
are replaced with GCs.



\subsection{Handling Refresh in Gain Cell Caches}

One of the biggest challenges in GC caches is the need to 
refresh. In addition to adding energy overheads, the cache is made unavailable 
for access during refresh operations, which adversely affects performance.
We use a staggered, concurrent refresh 
mechanism~\cite{DRAMwithConcurrentRefresh} to reduce unavailability 
of GC cache.

As explained earlier in this section, one way of the cache is 
mapped to a row in the GC sub-array. 
Refreshing one row of the sub-array takes 3~ns.
We refresh one row in a sub-array at a time and iterate over all the rows in a 
round-robin fashion in the course of the 1.12~ms refresh window, which is the DRT of an individual cell.

A refresh is done by reading the sub-array in the first 1.5~ns
of the refresh window. Data is written back to the row in the second half of 
the window. As a result, the sub-array is available for write in the 
first half and a read in the second half. This is made possible due to the 
presence of separate read and write bitlines. This optimization increases the 
availability of the sub-array and hence, the associated way -- it is now 
unavailable only for 1.5~ns every 4.375 $\mu$s. Since the tags are maintained in SRAM,
the cache can still be accessed to check for hits/misses.

We carry out a detailed analysis of performance and energy 
implications of this refresh policy, in Section~\ref{section:impactofrefresh}, 
and conclude that the performance overheads are minimal, since a tiny fraction (0.003\%) of cache accesses happen concurrently with refresh, leading to a \emph{worst-case} 1.7\% reduction in performance, as compared to the SRAM baseline. 


\section{Evaluation Methodology}
\label{section:experimental_methodology}

We evaluate our proposed architectures by using an 8-core system with
configuration listed in Table~\ref{table:systemconfig},
simulated using Sniper~\cite{Sniper}. This configuration is used as
the baseline for evaluation (ALL-SRAM). For iso-area GC caches
(-AREA), cache capacity is doubled by doubling
associativity. We test our proposals against 25 benchmarks from the SPEC CPU2017~\cite{sarabjeet2019_SPEC} and
PARSEC~\cite{PARSEC} suites and study energy and performance
implications in Sections \ref{section:energyanalysis} and \ref{section:performanceanalysis}, respectively. 
These workloads, listed in
Table~\ref{table:workloads}, are simulated for 2 billion instructions each, after a
500 million warmup period. Additionally, to include variations in the application behavior and test against multi-programmed
workloads, we divide the workloads in six sets, each consisting of 8 benchmarks, which represents: memory-intensive applications
(\textbf{MEM\_HIGH}), applications with average memory access (\textbf{MEM\_MED}), applications with sparse memory access
(\textbf{MEM\_LOW}), and three random mixes of 8-workloads (\textbf{mix1-3}). This classification is done based on memory
accesses per kilo instructions to caches - higher accesses means more memory intensive. Also, we create six homogeneous multi-programmed workloads (\textbf{8x*}) - 8 copies of the same benchmark, each per core.

\begin{table}[h]
\centering
\caption{System Configuration (ALL-SRAM)}
\label{table:systemconfig}
\scriptsize
\begin{tabular}{|c|c|} 
\hline
Processor & 8-core, 3.4GHz, x86\_64 ISA, 19-stage OOO \\ \hline 
Decode, Rename, Fetch Width & 4-7 fused, 4, 6 instructions per cycle \\ \hline 
Issue, Dispatch, Commit width & 4, 6, 4 fused $\mu$-ops per cycle \\ \hline 
ROB/Branch misprediction & 168 entries/8 cycles penalty \\ \hline 
L1-I/L1-D cache & 32 KB, 8-way \& 2 cycles. 64B line \\ \hline 
L2 cache & 256 KB, 8-way \& 5 cycles. 64B line \\ \hline 
L3 cache & Shared 8 MB, 16-way \& 10 cycles. 64B line \\ \hline 
Main Memory & \begin{tabular}[c]{@{}c@{}}4096MB DDR3, 100 ns access, \\ Read/Write Energy per 64B (nJ) = 41.6/54.4\end{tabular} \\ \hline 
\end{tabular}
\normalsize
\end{table}

\begin{table}[h]
\caption{Workloads}
\centering
\label{table:workloads}
\scriptsize

\begin{tabular}{|l|}
\hline
\textbf{Single-Programmed} \\ \hline \hline
\multicolumn{1}{|c|}{\textbf{SPEC CPU2017}}  \\ \hline
perlbench\_r, gcc\_r \\ 
bwaves\_r, mcf\_r \\ 
cactuBSSN\_r, parest\_r \\
povray\_r, lbm\_r  \\ 
omnetpp\_r, wrf\_r  \\ 
xalancbmk\_r, cam4\_r \\ 
deepsjeng\_r, imagick\_r \\
nab\_r, roms\_r, xz\_r  \\\hline
\multicolumn{1}{|c|}{\textbf{PARSEC}} \\ \hline
canneal, dedup \\ 
facesim, ferret \\ 
fluidanimate, freqmine  \\
raytrace, streamcluster \\ \hline
\end{tabular}
\quad
\begin{tabular}{|c|l|}
\hline
\multicolumn{2}{|c|}{\textbf{Multi-Programmed}} \\ \hline \hline
\multicolumn{1}{|c|}{\textbf{MEM\_HIGH}} & \multicolumn{1}{c|}{\textbf{MEM\_MED}} \\ \hline
\begin{tabular}[c]{@{}l@{}}cactuBSSN\_r, mcf\_r,\\ streamcluster, gcc\_r,\\ canneal, omnetpp\_r,\\ facesim, perlbench\_r\end{tabular} & \begin{tabular}[c]{@{}l@{}}xalancbmk\_r, ferret,\\ cam4\_r, bwaves\_r,\\ deepsjeng\_r, wrf\_r,\\ povray\_r, freqmine\end{tabular} \\ \hline
\multicolumn{1}{|c|}{\textbf{MEM\_LOW}} & \multicolumn{1}{c|}{\textbf{mix1}} \\ \hline
\begin{tabular}[c]{@{}l@{}}raytrace, parest\_r,\\ fluidanimate, nab\_r,\\ dedup, imagick\_r,\\ roms\_r, lbm\_r\end{tabular} & \begin{tabular}[c]{@{}l@{}}bwaves\_r, xz\_r,\\ wrf\_r, raytrace,\\ roms\_r, dedup,\\ lbm\_r, freqmine\end{tabular} \\ \hline
\multicolumn{1}{|c|}{\textbf{mix2}} & \multicolumn{1}{c|}{\textbf{mix3}} \\ \hline
\begin{tabular}[c]{@{}l@{}}perlbench\_r, ferret,\\ parest\_r, canneal,\\ omnetpp\_r, cam4\_r,\\ nab\_r, streamcluster\end{tabular} & \begin{tabular}[c]{@{}l@{}}mcf\_r, cactuBSSN\_r,\\ povray\_r, xalancbmk\_r\\ deepsjeng\_r, imagick\_r\\ facesim, fluidanimate\end{tabular} \\ \hline
\multicolumn{2}{|c|}{\textbf{8x*} - Running 8 copies of the same benchmark} \\ \hline
\end{tabular}
\normalsize
\end{table}

\begin{figure*}[h]
  \centering     
  \includegraphics[width=\linewidth]{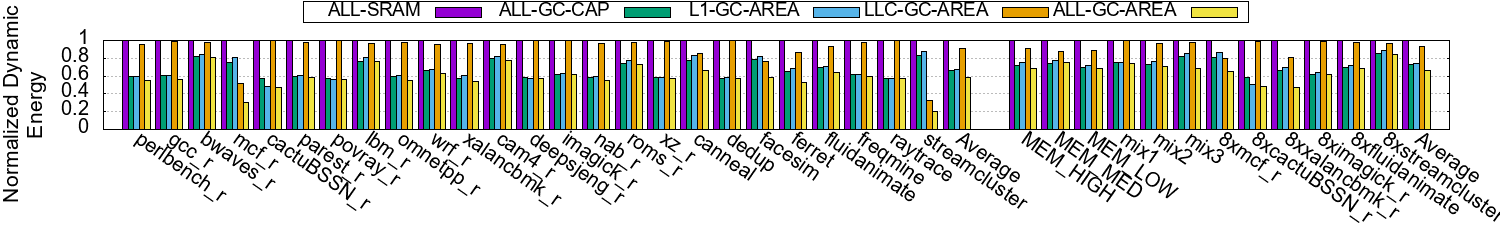}
  \caption{Dynamic energy (Cache + Main Memory) for single \& multi-programmed workloads. Normalized against ALL-SRAM.}
  \label{fig:Energy_x86}
\end{figure*}

\section{Evaluation of GC Caches}
\label{section:experimental_results}




In this section, we quantify the benefits of various GC based architectures and compare them with SRAM based
caches. The evaluation includes a study of memory subsystem
energy, performance, and the impact of refresh operations.
\subsection{Energy Analysis}
\label{section:energyanalysis}
While most emerging memory technologies exacerbate energy
requirements of the memory subsystem~\cite{SurveyofMemoryArchitectures,L3C_EmergingTechinCache,PCM_EvaluatingasMainMemoryAlt,STTRAM_EvaluatingasMainMemoryAlt}, GCs, on the contrary, provide significant energy
savings over SRAM. In comparison with the baseline SRAM, at the array level, proposed GC design consumes \textasciitilde 46\% less energy per read and 40-50\% less energy per write, as shown in Table~\ref{table:cachetiming}.

The ALL-SRAM case, where all levels of caches are assumed to be SRAM, is used as the baseline. We first study
ALL-GC-CAP, where we replace all SRAM caches with the same capacity GC caches (iso-capacity). Next, we evaluate iso-area GC caches with
double capacity.  
For this, we evaluate three configurations: (a)
L1-GC-AREA, where we replace L1 cache in ALL-SRAM with a double capacity GC cache. (b) LLC-GC-AREA, where we replace last-level
cache in ALL-SRAM with double capacity GC, and (c) ALL-GC-AREA,
where we replace \textit{all} levels of caches in ALL-SRAM with double capacity, iso-area GCs. 

Figure~\ref{fig:Energy_x86} compares the dynamic energy consumed by
the memory subsystem in proposed architectures, for both single and multi-programmed workloads. We calculate the dynamic energy consumption of caches and main memory by taking the product of the total number of accesses to each level with the energy consumption of per access using the per bit cache access energy (mentioned in Table~\ref{table:cachetiming}). Access energies of main memory are obtained from~\cite{MICRON} and are presented in
Table~\ref{table:systemconfig}.  
We observe that any cache hierarchy devised using GCs exhibits 
savings in dynamic energy.  In L1-GC-AREA, where only the L1 is
architected using proposed GCs, results in a 36\% reduction in dynamic
energy, averaged across all the single programmed benchmarks. The
 LLC-GC-AREA configuration, which replaces the SRAM LLC with a
 double capacity GC cache, also exhibits a 4\% average reduction 
 in dynamic  energy. Similar results are obtained for
 multi-programmed workloads as well. For the memory-intensive 
 mix (MEM-HIGH), the energy savings of L1-GC-AREA and LLC-GC-AREA 
 stand at 25\% and 9\%, respectively. 
 
 Finally, using GC for \textit{all} levels of the cache increases these gains tremendously. On average, across the single programmed
 workloads, ALL-GC-AREA achieves \textbf{42\%} (\textbf{34}\% in the case of multi-programmed) reduction in dynamic energy consumption as compared to the ALL-SRAM baseline. 
Even in cases where the area density benefits of proposed GC are not being utilized, i.e., in the sub-optimal configurations of ALL-GC-CAP, where all SRAM caches are replaced  with \textit{equal} capacity GC caches, we observe an average reduction in the dynamic energy of 34\% and 28\% for single and multi-programmed workloads respectively.
 



Compared to the baseline, applications like \textit{streamcluster}
and \textit{mcf\_r}, that have large number of memory accesses, achieve up to 80\% reduction in dynamic
energy for iso-area GC LLCs (LLC-GC-AREA). Increasing LLC capacity
allows the working set of these applications to reside in the cache,
reducing the number of off-chip accesses substantially (by \textasciitilde 99\%). Reduced off-chip accesses reduce the high off-chip
dynamic energy, resulting in massive energy savings.
Additionally, in a large, many-core CPU running at a low voltage, leakage from
on-chip caches contributes substantially to the chip's power
 draw~\cite{EDRAM_Refrint}. Proposed GC, with a large savings of
 99.3\% in leakage energy, as depicted in
 Figure~\ref{fig:leakage_power_comparison}, helps reduce these costs substantially.
 
\begin{figure*}[ht]
  \centering     
  \includegraphics[width=\linewidth]{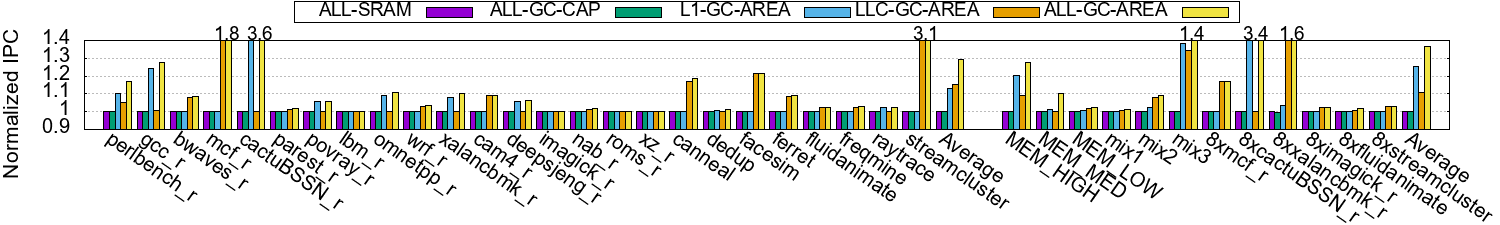}
  \caption{System performance for single \& multi-programmed (IPC is added across all cores) workloads. Normalized against ALL-SRAM.}
  \label{fig:IPC_x86}
\end{figure*}

\begin{figure*}[h]
  \centering
  \includegraphics[width=\linewidth]{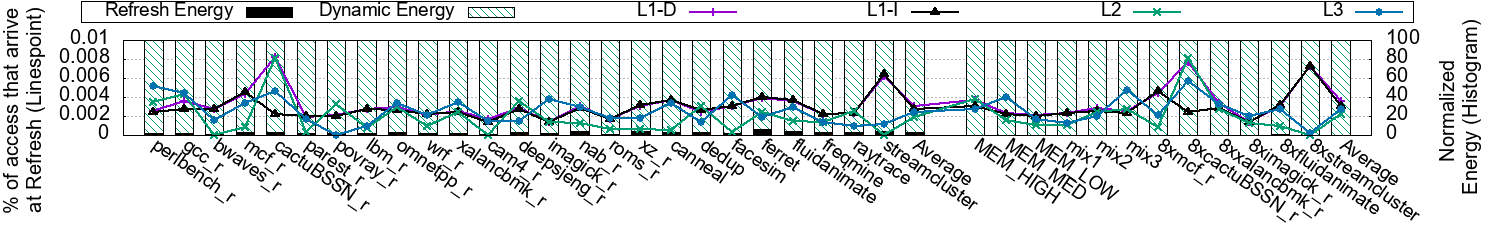}
  \caption{Percentage of accesses to the cache which arrive when the cacheline is being refreshed \textit{(y1-axis)}. Breakup of energy consumption of caches, normalized to total energy (Dynamic + Refresh) \textit{(y2-axis)}. Configuration used is ALL-GC-AREA.}
  \label{fig:Refresh_Impact}
\end{figure*}

\subsection{System Performance}
\label{section:performanceanalysis}
Figure~\ref{fig:IPC_x86} illustrates the system performance in terms
of instructions per cycle (IPC) for various proposals, using single and multi-programmed workloads. We observe that the performance difference between iso-capacity SRAM caches (ALL-SRAM) and GC caches (ALL-GC-CAP) is negligible - 0.1\% drop in IPC for GC caches on average, with respect to ALL-SRAM. All GC based iso-area caches (ALL-GC-AREA) exhibit \emph{performance gains} as compared to the baseline. Average performance increase of \textbf{29\%} and \textbf{36\%} is observed across single and multi-programmed workloads, respectively. In the case of multi-programmed workloads, memory-intensive workloads tend to benefit most. We observe a 27\% performance increase for MEM\_HIGH workloads mix, as compared to the baseline. 
We verified our results on an aggressive processor configuration, with better prefetcher, replacement policy and DRAM access latency~\cite{pref_GHB,pref_nextline,CRP_2}, and observed similar benefits ($<$4\% IPC drop from above reported benefits).
  
Applications like \textit{streamcluster}, \textit{canneal}, and \textit{mcf\_r} have working set sizes
that exceed the capacity of baseline SRAM caches. 
Hence, when the LLC size is doubled (LLC-GC-AREA), they see large performance improvements ($>$200\%) as working sets can reside on caches.
While, many applications like
\textit{cactuBSSN\_r} and \textit{gcc\_r} have working sets
that reside in the on-chip memory and leverage larger cache size to
fit working sets in the L1 cache. As a result, they achieve significant
performance improvements in L1-GC-AREA implementation (drop-in
accesses to next level caches by $>$90\%) but not in LLC-GC-AREA. On
average, L1-GC-AREA and LLC-GC-AREA achieve IPC improvements of 13\%
and 15\%, respectively.


\subsection{Impact of Refresh}
\label{section:impactofrefresh}
Regular GC based caches are required to refresh cells at regular intervals.
Unfortunately, this has adverse effects on both energy and performance. 1T1C eDRAM and traditional 2T, and 3T GCs can have huge refresh energy overheads, accounting for up to 
97\% of total LLC energy, as was observed in the experimental
results presented in Figure~\ref{fig:eDRAM_LLC}.

However, for caches designed using the proposed GC, owing to high
DRTs and staggered refresh mechanisms, we observe that the refresh energy consumption is minimal, assuming the most pessimistic scenarios. For experiments carried out with an all GC based cache subsystem (ALL-GC-AREA), which should exhibit the worst case refresh energy consumption profile, we observe that on average, across all single programmed benchmarks, refresh energy contributes \textbf{$<$3\%} (6\%, at max for \textit{ferret}) of the total energy consumption of all caches. We illustrate these observations in Figure~\ref{fig:Refresh_Impact}. For this worst case, we show that refresh energy has an insignificant contribution to the total energy, as depicted in the histogram on y2-axis.
 
Besides, the refresh operations do not affect performance adversely, as demonstrated from ALL-GC-CAP results from
Section~\ref{section:performanceanalysis}. This is evidenced by the
fact that caches spend only 0.008\% of the time on refresh, on average. Our experiments show that, on average, \textasciitilde 0.003\% of accesses to caches were made during refresh interval throughout the entire
simulation (Figure~\ref{fig:Refresh_Impact}, y1-axis).

\section{Asymmetric Writes}
\label{section:asymmetricwrites}
To read/write a value in a 6T SRAM cell, the bitlines first have to 
be precharged to V$_{DD}$. Then, wordlines are turned on to access the cell
value. On a read, both bitlines are precharged to high
(V$_{DD}$) while on a write, one bitline is driven to high and other to
low~\cite{weste,rabaey}. In proposed GC, we have separate bitlines for read 
(RBL) and write (WBL), as depicted in Figure~\ref{fig:schematic_GC}.
Due to this decoupling, there is no need to precharge the WBL to V$_{DD}$ 
before every access, unlike in  SRAM where the bit lines are pre-charged to $V_{DD}$ before every write. To perform a write, the WBL is precharged to high or low, 
depending on the value to be written. For instance, to write a $0$, WBL
will initially be connected to ground to establish the voltage
difference to drain the charge in the cell to $0$. Thus, if WBL was
already set to $0$, driving it again to $0$ would require no energy. Such cases
arise when the consecutive writes by WBL are the same, i.e., $0\rightarrow0$ or
$1\rightarrow1$ transitions. In these cases, the second write will have a
$0.48-0.67\times$ lower energy consumption as compared to a similar transition 
in 6T SRAM, as depicted in Table~\ref{table:cachetiming}.

\begin{figure*}[h]
  \centering
  \includegraphics[width=\linewidth]{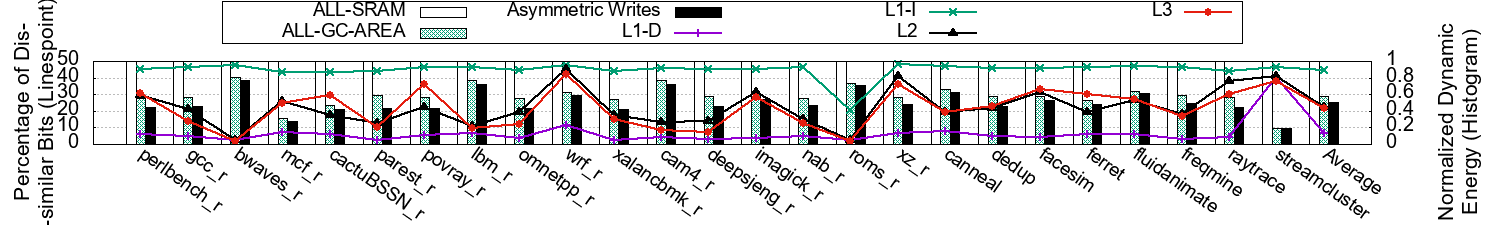}
  \caption{Dissimilar bits among consecutive writes, per cache level \textit{(y1-axis, \%)}. Dynamic Energy (Cache + Memory), normalized to ALL-SRAM \textit{(y2-axis)}.}
  \label{fig:asymmwrites}
\end{figure*}

\begin{figure}
  \centering
  \includegraphics[width=\linewidth]{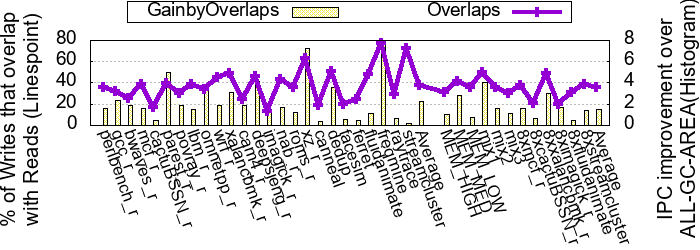}
  \caption{Percentage of writes that overlap with reads \textit{(y1-axis)}. Accordingly, the increase in IPC over ALL-GC-AREA \textit{(y2-axis).}}
  \label{fig:DecoupledRW_IPC}
\end{figure}


We quantify the overall dynamic energy savings due to such asymmetric
writes by calculating the number of dissimilar bits between two
consecutive writes. 
Dissimilar bit writes are serviced normally, while writes with similar
bits are serviced with reduced energy. The reduced energy parameters for
 caches, extracted from CACTI, are depicted in Table~\ref{table:cachetiming}. 
 We perform this experiment with ALL-GC-AREA configuration 
 and present our results in Figure~\ref{fig:asymmwrites}. On y1-axis,
 we present the ratio of dissimilar bit writes to total bit writes, for each 
 cache. Lower ratio implies that a lot of the data being written is 
 similar to the value of the WBL, and hence can be written with lower write energy. Accordingly,
 we calculate the overall dynamic energy consumption (cache + main memory) and compare it with baseline ALL-SRAM and ALL-GC-AREA proposals on y2-axis.
We observe that most  bits -- 76\%, averaged across all levels of caches, in write data are similar to the write bitline's value. 
SPEC CPU2017 workloads rarely have writes which are larger than 8Bytes~\cite{sarabjeet2019_SPEC}.
The rest of the cacheline is re-written with the same value. This is true for all data caches, with L1D exhibiting as
high as 94\% similarity, averaged across all benchmarks.
Even in cases, where the initial access was a miss, and the existing cacheline has to be replaced with a new one being brought in, we 
observe significant data similarity between the new and the old lines.
For instance, L1 I-cache, which only experiences writes as
insertions of new cachelines, observes a 55\% data similarity between the old
and new lines. As a result, we observe a \textbf{13\%} reduction 
in dynamic energy consumption, with respect to ALL-GC-AREA, and \textbf{50\%} compared to baseline SRAM-based cache subsystem (ALL-SRAM). Applications with higher write ratios~\cite{sarabjeet2019_SPEC} tend to save more energy, for example, \textit{parest\_r}, \textit{omnetpp\_r}, \textit{xalancbmk\_r}, and \textit{povray\_r}. Therefore, we show that, because of its inherent structure, GC can take advantage of write similarity in data to further save on dynamic energy.

Additionally, due to decoupled bitlines, reads and writes to the same sub-array can be done in parallel. We use this property to overlap writes with 
simultaneously occurring reads to the same sub-array. We note that 40\% of all writes could be overlapped with \textit{some} reads, represented by line-graph \textit{Overlaps} in Figure~\ref{fig:DecoupledRW_IPC}. Most of the overlaps happen in L1-D cache, as L2 and L3 caches have more sub-arrays and experience a smaller number of writes than L1. By hiding the latency of these writes, we observe a \textasciitilde 2\% increase in IPC compared to ALL-GC-AREA, presented by \textit{GainbyOverlaps} in Figure~\ref{fig:DecoupledRW_IPC}.
 To further take advantage of decoupled bitlines, smart cacheline placement policies or buffering can be used, which can potentially increase the number of overlaps. However, we do not pursue these optimizations 
 due to the lack of substantial returns on either performance or energy.

\section{Hybrid Cache Hierarchy}
\label{section:TechComparison}
 
 \begin{table}
\caption{LLC parameters for different technologies}\label{table:LLCparameters}
\centering
\scriptsize
\begin{tabular}{|l|c|c|c|}
\hline
& \textbf{eDRAM}
& \textbf{STTRAM} & \textbf{Hybrid} (8MB \textbf{GC}\\ 
& 32MB & 32MB &+16MB \textbf{STTRAM)}\\ \hline
Read Latency (ns) \textbf{(cycles)}& 5.15 \textbf{(18)} & 26 \textbf{(89)} & 2.93 \textbf{(10)}, 26 \textbf{(89)} \\ 
\hline
Write Latency (ns) \textbf{(cycles)}& 5.15 \textbf{(18)} & 60 \textbf{(204)} & 2.93 \textbf{(10)}, 60 \textbf{(204)} \\ \hline
Read/\textbf{Write} Energy / bit (pJ) & 5.2/\textbf{6.12} & 5.35/\textbf{7.85} & 3.81/\textbf{5.52}, 5.35/\textbf{7.85} \\ \hline
Refresh Interval/\textbf{Period}& 0.02ms/\textbf{4ns} & -/- & 1.12ms/\textbf{1.5ns}, -/\textbf{-} \\ \hline
Refresh Energy/bit (pJ) & 3.5 & - & 1.87/ - \\ \hline
\end{tabular}
\normalsize
\end{table}

\begin{figure*}[h]
  \centering     
  \includegraphics[width=\textwidth]{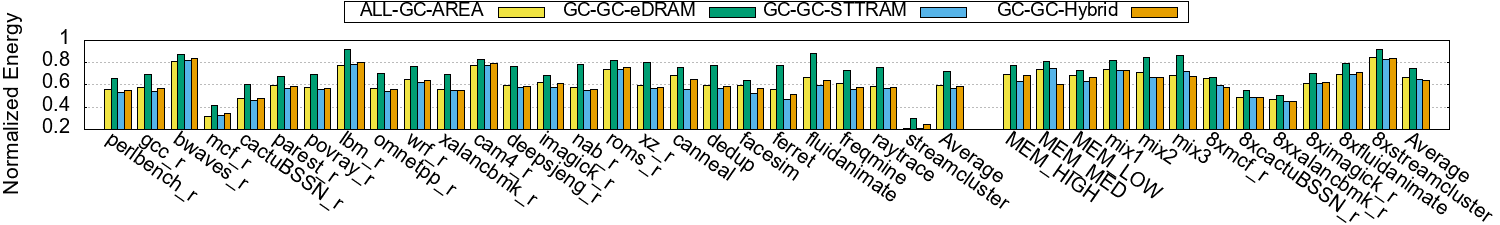}
  \caption{Dynamic + Refresh Energy for single \& multi-programmed Workloads. Normalized against ALL-SRAM.}
  \label{fig:Energy_Technologies}
\end{figure*}

\begin{figure*}[h]
  \centering     
  \includegraphics[width=\textwidth]{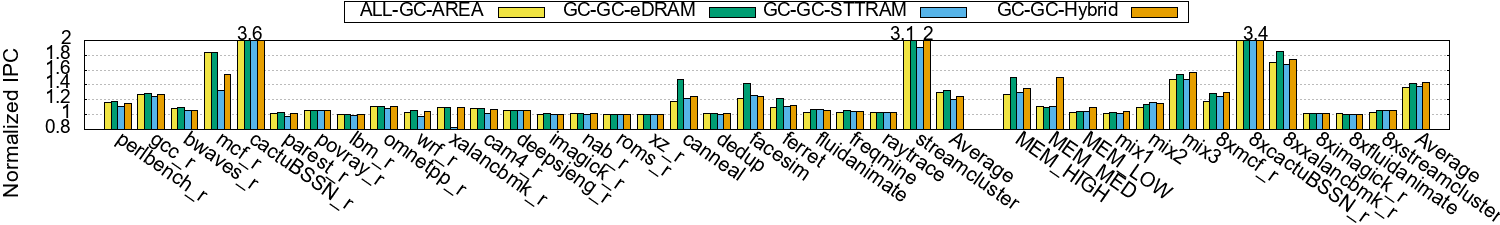}
  \caption{System performance for single \& multi-programmed (IPC is added across all cores) workloads. Normalized against ALL-SRAM.}
  \label{fig:IPC_Technologies}
\end{figure*}



A growing body of research has proposed either eDRAM or STT-RAM as a replacement for LLCs (\cite{STT_DASCA,STT_AOS,STT_khoshavi2016read,STT_rasquinha2010energy,STTRAM_Cache_TradeoffwithNonVolatility,STTRAM_Cache_3D,STTRAM_Cache_LLC,mram1,mram2,EDRAM_Refrint,EDRAM_Mosaic,L3C_EmergingTechinCache}). In this section, we build on prior work to evaluate hybrid cache
hierarchies, in an effort to build efficient SRAM ``free'' on-chip caches. 

First, we compare proposed GC based caches with other, state-of-the-art memory technologies. We consider the architectures, where L1 and L2 caches are kept as GC and use either eDRAM or STT-RAM in LLC, namely ``\textit{GC-GC-eDRAM}'' and ``\textit{GC-GC-STTRAM}'' respectively. STT-RAM parameters were taken from~\cite{mram2}, while parameters for eDRAM are obtained from CACTI simulations, and are listed in Table~\ref{table:LLCparameters}. In our experiments, we consider state-of-the-art refresh-optimized eDRAM~\cite{EDRAM_Mosaic} which achieves \textasciitilde 20$\times$ reductions in the number of refreshes over regular eDRAM at 2-3\% area overhead.

We compare these technologies with \textit{ALL-GC-AREA} and present energy 
and performance comparisons in Figures~\ref{fig:Energy_Technologies} 
and~\ref{fig:IPC_Technologies}, respectively.
These experiments provide several interesting results.
(a) eDRAM based LLC has 2$\times$ density benefits over GC, which helps it achieve 2.3\% (4\% for multi-core) improvement in IPC over ALL-GC-AREA, which makes a case for an eDRAM-based LLC. However, eDRAM has large refresh overheads. Even the refresh-optimized eDRAM~\cite{EDRAM_Mosaic} (GC-GC-eDRAM) results in 21\% (12\% for multi-core) more total energy consumption 
than ALL-GC-AREA. Traditional eDRAM results in much worse energy 
overheads -- 6.8$\times$ higher energy consumption as compared to ALL-GC-AREA.

STT-RAM, with the same 2$\times$ density benefits over GC, suffers from much longer access latencies, which have been enumerated in Table~\ref{table:technologycomparison}.
As a result, configurations with STT-RAM LLC experienced 
a performance drop of 7\% with respect to ALL-GC-AREA, even though the number of off-chip requests actually dropped significantly by 13\%. However, in realistic cases (multi-core runs) that take advantage of larger LLC, GC-GC-STTRAM performs slightly better than ALL-GC-AREA. Consequently, with smaller off-chip accesses, energy consumption reduces by 5\%, compared to ALL-GC-AREA, concluding that STT-RAM LLC would be a better design.


\begin{figure}
  \centering
 \includegraphics[width=\linewidth]{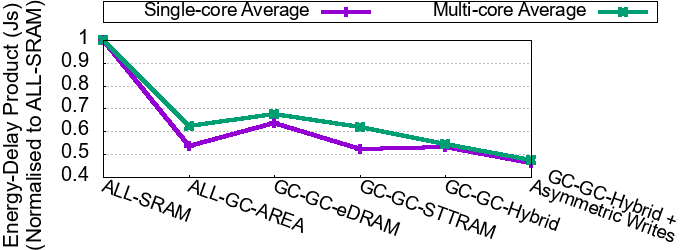}
  \caption{Energy Delay Product (Js) of proposals, normalized to ALL-SRAM.}
  \label{fig:EDP_Avg}
\end{figure}

In an effort to get the best of all worlds: utilize higher 
density of STT-RAM, and low latency of GC, we propose 
\textit{hybrid LLC} designs of GC 
and STT-RAM. We selected STT-RAM to avoid the high energy overheads imposed by eDRAM. We carried out a design space exploration for the optimal size partitioning between GC and STT-RAM, while maintaining the same area budget as 
SRAM, and found that equal-area (8~MB GC, 16~MB STT-RAM) distribution results 
in the sweet spot of high performance and low energy. The parameters of this organization are listed in Table~\ref{table:LLCparameters}.
The ways of each set of the hybrid cache are split between GC and STT-RAM cachelines, in the ratio of capacity. On a cache lookup, tags of both GC and STT-RAM ways are read and compared. If there is a hit in one of the GC ways, a read or write is carried out. On a miss in GC ways, but a hit an STT-RAM way, the cacheline is moved to the LRU position in the GC ways. Since GC ways tend to have ``hot'' data, in order to exploit temporal locality, the evicted cacheline from GC way is moved to the LRU position of the STT-RAM ways. In case of a miss in both GC and STT-RAM ways, the cacheline is fetched from the next level and placed in the LRU position of STT-RAM ways. 
The proposed hybrid cache architecture can be further optimized via novel replacement policies and prefetchers, which we leave for future work~\cite{PHC,Benzene}.
With L1 and L2 
cache as GC, and a hybrid LLC, we evaluate the architecture, 
results of which are compiled in orange bars of 
Figures~\ref{fig:Energy_Technologies} \& \ref{fig:IPC_Technologies}, under ``\textit{GC-GC-Hybrid}''.

As expected, the energy consumption of hybrid design is very close to that of  ALL-GC-AREA: within 2\%, on average, across both single and multi-core benchmarks. More importantly, the increased LLC accesses, due to larger cache, are performed with lower latencies of GC. As a result, we observe 5\% 
improvement in IPC over ALL-GC-AREA baseline, averaged across multi-core 
simulations. Compared to the traditional SRAM hierarchy, our proposal shows 
\textbf{24\%} better performance with \textbf{42\%} less energy consumption 
(\textbf{43}\% better with \textbf{36}\% less energy, in case of multi-core). 
These designs can be further optimized by exploiting decoupled bitline optimizations proposed previously, resulting in an extra 13\% savings in energy, as discussed in Section~\ref{section:asymmetricwrites}, leading to overall \textbf{50\%} (\textbf{44\%} in case of multi-core) saving in energy as compared to ALL-SRAM. In conclusion, while GC works best for L1 and L2 caches, real environments require large-capacity LLC with low latency, which can be addressed with our proposed 
GC-STTRAM hybrid LLC design. In the hybrid design, we move the recently
accessed cachelines to the GC part of hybrid LLC, thus serving them with lower latency, if locality exists. 

In summary, we present Energy-Delay Product (EDP) results of various architectures in Figure~\ref{fig:EDP_Avg}. As can be
observed, for both single and multi-programmed workloads, \textit{any} GC 
based hierarchy does better than the baseline SRAM one. The most favorable design point is obtained by utilizing the  
benefits of asymmetric write - optimized GC caches at all levels, 
and a hydrid STT-RAM - GC LLC. This architecture achieves an EDP 
which is \textbf{0.46}$\times$ of the baseline SRAM one.

\begin{figure*}[h]
  \centering
  \includegraphics[width=\textwidth]{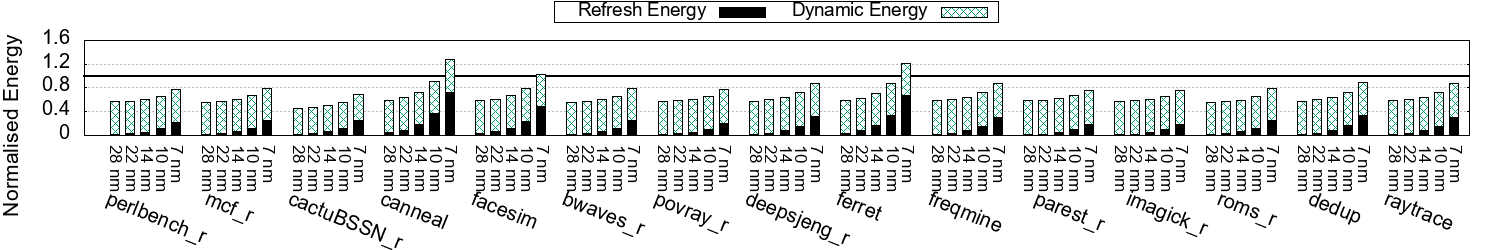}
  \caption{Energy breakdown (Cache) of ALL-GC-AREA at various technology nodes. Normalized to ALL-SRAM dynamic energy at same technology node.}
  \label{fig:TechScaling}
\end{figure*}

\begin{figure}
  \centering
  \includegraphics[width=0.5\textwidth]{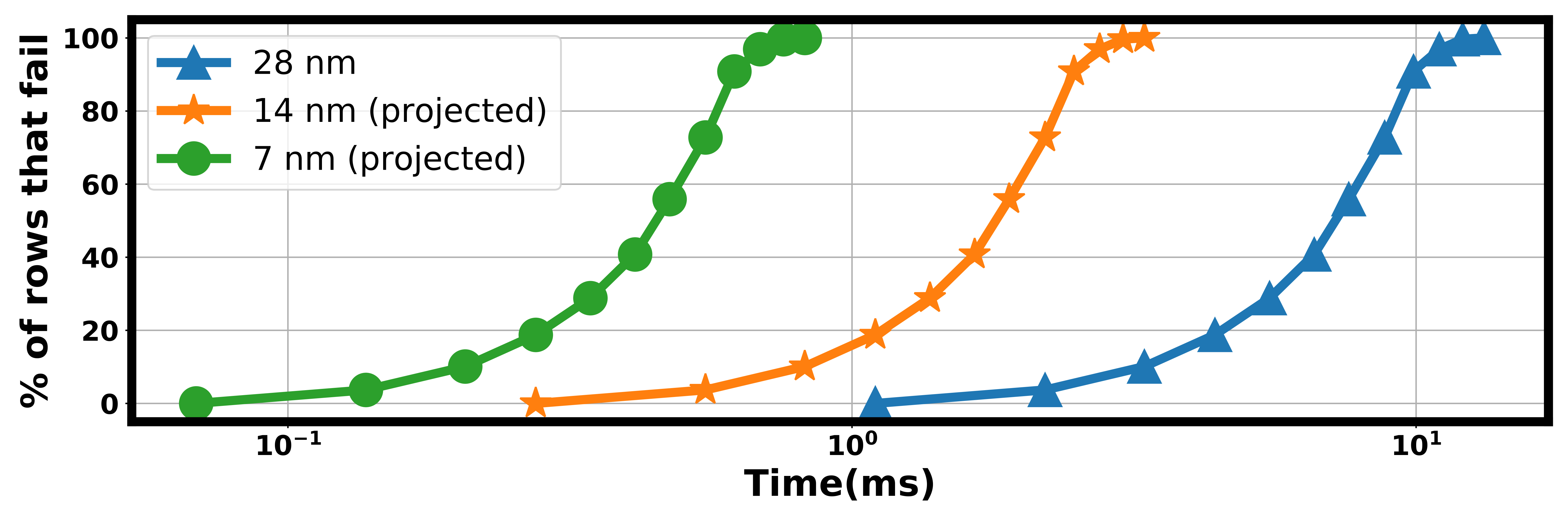}
  \caption{Data Retention Time for different rows, that decides refresh interval.}
  \label{fig:RAIDR}
\end{figure}

\section{Scalability of Proposed Caches}
\label{section:ScalingGC}

As the technology scales down, transistors become leakier with smaller storage capacitances, resulting in smaller DRTs, exacerbating the refresh problem. 
This means that at lower technology nodes, GC based caches, due to a large 
number of refreshes, could perform worse than traditional
SRAM based caches. To understand scalability characteristics of
proposed GCs, we carry out experiments to characterize the energy consumption of proposed GC with technology scaling.

We carry out our analysis for 28, 22, 14, 10, and 7 nm technology nodes. Generally, leakage current and device capacitance are inversely proportional to technology node~\cite{ITRS,weste}. So, scaling down to the next
technology node would result in DRT reduction by \textasciitilde $50$\%.
Additionally, cell access energies also
decrease by \textasciitilde $50$\% as we move to lower technology~\cite{weste}. Considering these trends, we simulate ALL-SRAM and ALL-GC-AREA 
configurations and calculate dynamic and refresh energy consumptions. 
At each technology node, we
normalize ALL-GC-AREA's energy consumption (dynamic + refresh) compared to ALL-SRAM's energy consumption (dynamic) and
present the results in Figure~\ref{fig:TechScaling}. Due to
space constraints, we present only 15 benchmarks, 5 each from
different memory intensity groups listed in Table~\ref{table:workloads}. 
The first five benchmarks are from MEM\_HIGH, followed by MEM\_MED and MEM\_LOW.

We observe that as we move to lower technology nodes, the
contribution of refresh energy \textit{increases considerably}. At 
7~nm, the energy consumption of GC based caches (ALL-GC-AREA)
is almost comparable to SRAM based caches (ALL-SRAM). Therefore,
at 7 nm or lower, proposed GC, due to refresh, can perform worse than SRAM. Many techniques~\cite{EDRAM_Refrint,ReducingCachePowerwithECC,EDRAM_Mosaic,EDRAM_RANA} have been explored to reduce refreshes if it
becomes a problem. We propose to reduce the refresh energy by increasing the back-gate bias voltage. 
Figure~\ref{fig:Id_Vgs_Backgate_characterics} shows reduction
in leakage current with increasing back-gate bias voltage in a negative direction. This increases the cell's DRT, which decreases refresh frequency and hence, energy. 
In Figure~\ref{fig:RAIDR}, for three technology nodes, we show
 that the actual DRT of rows can be much higher than the worst case,
 for which we have to design refresh mechanisms.
A potential solution to avoid that could be to use
architectural solutions like RAIDR~\cite{RAIDR} and 
implement separate refresh intervals for different
rows based on their actual DRTs. This can be achieved by 
dividing the rows into DRT bins
and applying a different refresh interval for each bin, leading to a reduction in the number of refresh operations.

\begin{figure*}
  \centering
  \includegraphics[width=\textwidth]{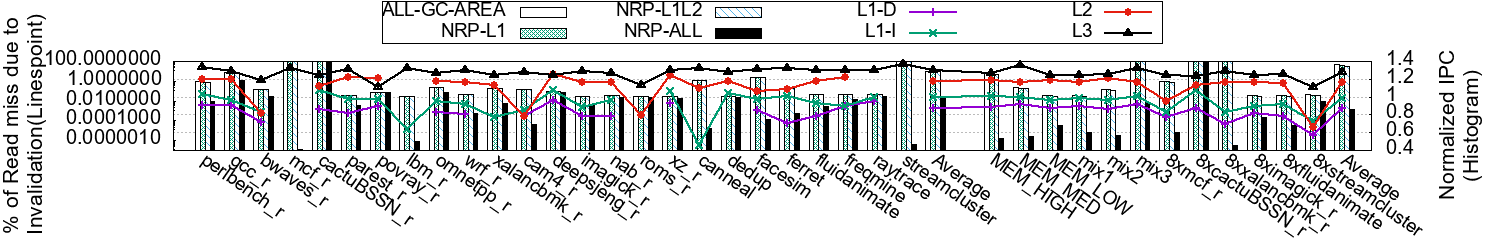}
  \caption{Percentage of Reads that miss due to No Refresh Policy (NRP) invalidations \textit{(y1-axis)}. Performance of NRP, normalized against ALL-SRAM \textit{(y2-axis)}.}
  \label{fig:NRP_IPC}
\end{figure*}

\subsection{Refresh Free Hybrid Cache Hierarchy}
Another key solution which reduces the effect of refresh is a \textit{No Refresh Policy (NRP)}, where rather than refreshing a cacheline, we invalidate the data if the line has not been 
touched after a refresh operation has been performed, 
but before it reaches its DRT. The key observation here is that 
if a row is accessed, the countdown for its DRT is reset, starting
the refresh window from that point. 
NRP can further be optimized by taking advantage of the high
number of writes in caches to reduce such invalidations (similar to~\cite{SmartRefresh}).
To invalidate GC cachelines, a dedicated, on-chip refresh controller, which generates a pulse at given intervals would be required~\cite{EDRAM_Refrint,MicroRefresh}. We propose to maintain a 5-bit (programmable) saturating counter for every cacheline, which is incremented every $1/32^{th}$ epoch of the refresh interval. Once the counter saturates, the cacheline is invalidated. If the cacheline was dirty, it is written back to the next level of the hierarchy. In case of a write to the cacheline, the counter is reset to 0.
We implement this policy at 28nm and calculate the number of times a cacheline was invalidated due to DRT expiration and present
the percentage of cache read misses due to such invalidations
 on the y1-axis of Figure~\ref{fig:NRP_IPC} and the performance impact
 of these invalidations on the y2-axis. We observe
 that NRP in L1 and L2 caches does not cause many misses ($<$1\%), and exhibits similar performance as compared to
 GC with refreshes - on average, $<$0.1\% and \textasciitilde 1\% drop in IPC, for single and multi-programmed workloads respectively.
 However, extending NRP to LLC generates \textasciitilde 12\% new
 misses due to invalidations. It, therefore, results in 23\% (36\% for multi-programmed workloads) drop in IPC (as seen from Figure~\ref{fig:NRP_IPC}) as a miss in LLC results
 in a request to the main memory. 
 Also, while invalidating, we writeback dirty cachelines. This has negligible ($<$10MB/s) bandwidth impact on L1 and L2 bandwidth, but results in an average memory bandwidth usage of \textasciitilde 110MB/s across multi-core workloads, which is small but may be unacceptable in many cases. Although NRP doesn't generate many new writebacks because it preemptively invalidates cachelines which would otherwise have been dirty evictions.
 NRP can
 altogether remove refreshes, and hence refresh energy, from L1 and L2 caches, while L3 cache would still need to be periodically refreshed. However, implementing L3 with STT-RAM does not need not refresh. As a result, we conclude that a hybrid cache hierarchy, where the L1 \& L2 comprise of proposed
 GC, and LLC is made from STTRAM-GC hybrid is the lowest energy, highest performance on-chip cache hierarchy with refresh-free L1 and L2 caches.

\section{Related Work}
\label{section:relatedwork}

\textbf{\textit{Emerging Memory Technologies for Caches:}}
Due to scalability and energy issues of traditional SRAM, several studies have been carried out to evaluate emerging memory technologies for caches. STT-RAM, owing to its low leakage energy and density benefits, has been viewed as a promising candidate. However, it suffers from inherent weaknesses - high write latency and write energy. There have been many proposals to alleviate these shortcomings~\cite{STTRAM1,STT_imani2016low,STT_khoshavi2016read,STT_samavatian2014efficient,STT_wang2014adaptive,STT_EarlyWriteTermination,STTRAM_Cache_TradeoffwithNonVolatility,STT_rasquinha2010energy,STT_AOS,STTRAM_Cache_3D,STT_DASCA}.

Another potential replacement for SRAM are eDRAMs, which offer high density, low leakage, similar access latencies, and low dynamic energies. However, eDRAM requires refresh operations to preserve data integrity~\cite{arch2}. As cache 
size increases, each refresh requires more energy, and more lines need to be 
refreshed; thus, refresh can potentially become the main source of eDRAM power 
dissipation. Many studies have been carried out to amortize this effect~\cite{ReducingCachePowerwithECC,EDRAM_RANA,EDRAM_Refrint,L3C_EmergingTechinCache,HybridCache_Disparate,PCM_Cache_Hybrid,MicroRefresh}. For instance, \cite{EDRAM_Mosaic} 
showed that the DRTs of cells in large eDRAM modules exhibit spatial 
correlations, and exploit this behavior to reduce refresh energy. In contrast, 
the proposed GC already has insignificant refresh overheads. 

\textbf{\textit{Gain Cell:}}
Many circuit and array level architectures have been proposed for GCs~\cite{capacitorless,capacitorless2,capcitorless4,capacitorless5,2Tcache,3Tcache,4Tcache,gaincellnew,gaincellnew2,gaincellnew3}. The biggest advantage of the GC is its logic compatible fabrication process - only transistors are used for designing the cell and, therefore, can use the same fabrication process as the processor. Transistor's parasitic 
capacitor is used to store the data. Due to low storing capacitance and high
leakage of the junctions in bulk-MOSFET \cite{weste}, DRT of GC is tiny. To improve DRT, Robert et al.\cite{4Tcache} propose using FDSOI-MOSFET instead of bulk-MOSFET or FinFET, which has reduced leakage current as the junctions are isolated in FDSOI through the oxide \cite{fdsoi1}. The other added advantage of the FDSOI device is its back-gate bias feature \cite{powergating}. While body-bias can also be applied in bulk devices to improve the DRT \cite{bodyimpact}, it has limited breakdown voltage and can only be applied in p-type MOSFETs \cite{weste}. In contrast, a large back-gate bias can be applied in FDSOI technology. 

In the FDSOI device, threshold voltage and leakage of the device can be changed dynamically with the help of the back-gate bias. Many designs \cite{back-gate1,back-gate2,back-gate3,back-gate4} have been proposed by exploiting the back-gate bias feature of the FDSOI. \cite{powergating} designed the power gating circuit with the help of the back-gate bias feature of the FDSOI device. Clerc et al.~\cite{multiplier} designed the robust multiplier and DC-to-DC converter by exploiting the back-gate bias feature. In \cite{4Tcache}, GC is implemented using FDSOI-MOSFET for its low leakage feature and uses 2 additional transistors to improve DRT at the cost of the area.


\textit{\textbf{Architectural optimizations for SRAM caches:}}
 Many schemes have been looked into to reduce SRAM's leakage energy~\cite{SRAM_CacheDecay,SRAM_Drowsy}.
To catch up with the increasing demands for large capacity caches, numerous cache compression techniques have been looked into~\cite{compression_survey,compression_Panda,compression_HyComp}. 
Also, there has been a significant amount of work on achieving higher performance in caches. Studies have been carried out for efficient cache replacement and cache management policies~\cite{CRP_1,CRP_2,CRP_3,CRP_4,CRP_5}, dead block predictions~\cite{DBP_1,DBP_2,DBP_3}, or exploiting the differences between reads and writes in caches~\cite{EW_1,EW_2}.
\section{Conclusion}
\label{section:conclusion}

In this work, we propose a novel FDSOI based 2T Gain Cell (GC) as a promising candidate for use in all levels of on-chip caches. The proposed GC has better energy efficiency, a much smaller area footprint, and better scalability as compared to 6T SRAM. Using the back-gate bias feature of FDSOI, we improve GC's data retention time, minimizing contribution of refresh to the overall energy consumption of the cache hierarchy.

We evaluate various architectural implementations of GC, for all levels of on-chip caches and demonstrate that GC based caches substantially reduce the dynamic energy consumption of memory subsystem as compared to traditional SRAM 
caches. We exploit the inherent capabilities of the proposed GC,
including decoupled read and write bitlines, to further reduce dynamic 
energy. 
Next, we demonstrate that a no-refresh policy, where GC based cache lines 
are invalidated than refreshed, can be a viable implementation
for removing refreshes in GC caches closer to the CPU.
Finally, we show that proposed GC, in conjunction with 
emerging memory technologies like STT-RAM can be used to architect  SRAM-free cache hierarchies with a much superior energy-delay product
as compared to SRAM caches.

\bibliographystyle{ACM-Reference-Format}
\bibliography{main}

\end{document}